\documentclass{webofc}
\usepackage[varg]{txfonts}   
\woctitle{Physics at the Magnetospheric Boundary}

\newcommand{\ba}{\begin{eqnarray}}
\newcommand{\ea}{\end{eqnarray}}
\newcommand{\be}{\begin{equation}}
\newcommand{\ee}{\end{equation}}
\def\go{\mathrel{\raise.3ex\hbox{$>$}\mkern-14mu
             \lower0.6ex\hbox{$\sim$}}}
\def\lo{\mathrel{\raise.3ex\hbox{$<$}\mkern-14mu
             \lower0.6ex\hbox{$\sim$}}}
\newcommand{\hatl}{\hat{\mbox{\boldmath $l$}}}
\def\bOmega{{\mbox{\boldmath $\Omega$}}}

\begin{document}
\title{Theory of Disk Accretion onto Magnetic Stars}
%
%

\author{Dong Lai\inst{1}\fnsep\thanks{\email{dong@astro.cornell.edu}}
}

\institute{Department of Astronomy, Cornell University, Ithaca, NY 14853, USA
}

\abstract{%
Disk accretion onto magnetic stars occurs in a variety of systems,
including accreting neutron stars (with both high and low magnetic
fields), white dwarfs, and protostars. We review some of the key
physical processes in magnetosphere-disk interaction, highlighting
the theoretical uncertainties. We also discuss some applications
to the observations of accreting neutron star and protostellar systems, as well as
possible connections to protoplanetary disks and exoplanets.
}
\maketitle

\section{Introduction}
\label{intro}

Disk accretion onto magnetic central objects occurs in a variety of
astrophysical contexts, ranging from classical T Tauri stars (e.g.,
Bouvier et al.~2007) and cataclysmic variables (intermediate polars;
e.g. Warner 2004), to accretion-powered X-ray pulsars (e.g. Lewin \&
van der Klis 2006). Such accretion has been studied extensively since
the discovery of high-mass accreting X-ray pulsars in the 1970s (e.g.,
Pringle \& Rees 1972). The basic picture of disk-magnetosphere
interaction is well known. The stellar magnetic field disrupts the
accretion flow at the magnetospheric boundary and funnels the plasma
onto the polar caps of the star or ejects it to infinity (e.g.,
Pringle \& Rees 1972).  The magnetosphere boundary is located where
the magnetic and plasma stresses balance,
\be
r_m =\xi \left({\mu^4\over GM\dot M^2}\right)^{1/7},                           
\ee
where $M$ and $\mu$ are the mass and magnetic moment of the central
object, $\dot M$ is the mass accretion rate and $\xi$ is a
dimensionless constant of order 0.5-1.
Roughly speaking, the funnel flow occurs when $r_m$ is less than the corotation
radius $r_c$ (where the disk rotates at the same rate as the
star). For $r_m\go r_c$, centrifugal forces may lead to ejection of
the accreting matter (``propeller'' effect; Illarionov \& Sunyaev 1975).

However, beyond this simple ``Astro-101'' picture (with the estimate of
$r_m$ given above), the situations are far more complicated.  Over the last
several decades, numerous theoretical studies have been devoted to
understanding the interaction between accretion disks and magnetized
stars. Many different models have been developed.  Some examples are
(see, e.g., Ghosh \& Lamb 1979; Aly 1980; Lipunov \& Shakura 1980;
Anzer \& B\"orner 1980,1983; Arons 1987,1993; Wang 1987,1995; Aly \&
Kuijpers 1990; Spruit \& Taam 1990,1993; K\"onigl 1991; Shu et al.~1994,2000; van
Ballegooijen 1994; Lovelace et al.~1995,1999; Li et al.~1996;
Wickramasinge \& R\"udiger 1996; Campbell 1997; Lai 1998,1999; Terquem
\& Papaloizou 2000; Shirakawa \& Lai 2002a,b; Pfeiffer \& Lai 2004;
Uzdensky et al.~2002, Uzdensky 2004; Matt \& Pudritz 2005; D'Angelo \&
Spruit 2010):

$\bullet$ Ghosh \& Lamb (1979) assumed that the stellar fields invade
the disk over a large range of radii. Differential rotation between
the star and the disk will cause the field lines to wind up. They
argued that a steady state can be reached, where the winding-up of the
field lines is balanced by dissipation, so that the field lines slip
across the disk.  A problem with the GL picture is that in order to
avoid too much winding, they need a very large (and unrealistic) 
magnetic diffusivity.

$\bullet$ In fact, the accretion disk is a very good conductor. So it
is possible that the stellar magnetic field cannot penetrate the disk
except near the magnetosphere boundary. This kind of models were first
explored by Arons, McKee \& Pudritz (see Arons 1986).  The field lines
are strongly pinched and a centrifugally driven wind may come out from
the magnetosphere boundary.  These ideas were worked out and expanded
in more details in the so-called x-wind model of Shu et al.~(1994),
which has a rich set of phenomenology and applications.

$\bullet$ Lovelace et al.~(1995) considered a model where the field
lines connecting the star and disk becomes open because of the winding-up.
So there is an inner magnetosphere corotating with the star but also
an open region where wind may come out.

$\bullet$ Matt \& Pudritz (2005) developed a model where the possible
role of stellar wind driven by accretion was emphasized..

\begin{figure}
\centering
\includegraphics[width=12.5cm,clip]{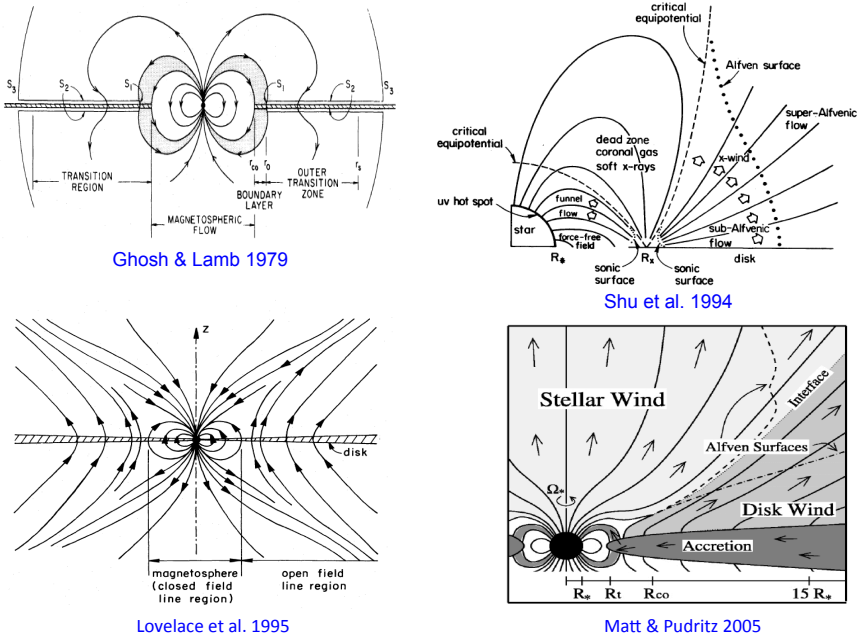}
\vskip -0.8truecm
\caption{Different theoretical models of magnetosphere-disk interaction (see text for 
discussion).}
\label{f1}       
\end{figure}

In parallel to these theoretical studies, since 1990s there have been
many numerical simulations, with increasing sophistication, on this
problem (Hayashi et al.~1996,2000; Miller \& Stone 1997; Goodson et
al.~1997; Fendt \& Elstner 2000; Matt et al.~2002; Romanova et
al.~2003,2009). These simulations are playing an important role in
elucidating the physics of magnetosphere-disk interaction in various
astrophysical situations (see papers by Romanova and by Zanni in this
volume).

In this paper, we will not discuss any particular models or
simulations, which are covered by others. Instead, we will focus on
several uncertain issues in the theory of magnetosphere-disk
interaction, and will also discuss their possible applications and
relevance in various astrophysical situations.

\section{Magnetospheric Boundary and Disk Inner Radius}
\label{sec-1}

Let us first consider the problem of dipole magnetic field of a star
being invaded by a conducting disk (Aly 1980). Note that an accretion disk is
indeed a good conductor. MRI-driven turbulence gives rise to 
both viscosity $\nu$ and magnetic diffusivity $\eta$, with the ratio
(called magnetic Prantl number) $\nu/\eta$ of order a few 
(see the simulation by Lesur \& Longaretti 2009). 
With the $\alpha$-parameterization, $\eta\sim\nu=\alpha H c_s$, the diffusion time
of magnetic field across the disk is $t_{\rm diff}\sim H^2/\eta\sim 1/(\alpha
\Omega)$ (where $\Omega$ is the disk rotation velocity), much longer than the
disk dynamical time. 

For a perfectly conducting disk/plate (with inner radius $r_{\rm in}$
surrounding a dipole, the exact solution (Aly 1980) shows that the vertical
field at the disk inner edge is $B_z(r_{\rm in})\simeq B_\star (r_{\rm in})
(r_{\rm in}/H)^{1/2}$, where $B_\star(r_{\rm in})\sim \mu/r_{\rm in}^3$ is the
dipole field at $r=r_{\rm in}$ when the disk is absent.

Now, at the inner edge of the disk, various instabilities (including
magnetic Kelvin-Helmholtz and Rayleigh-Taylor, and reconnection) can develop,
leading to the formation of a boundary layer. The stellar dipole field
penetrates this magnetospheric boundary layer and gets twisted to produce
$B_{\phi\pm}=\mp\zeta B_z$, with $\zeta\sim 1$ (the upper/lower sign specifies
the field above/below the disk). In the boundary layer ($r_m<r<r_m+\Delta r_m$),
the plasma angular velocity transitions from the Keplerian rate $\Omega_K$ to the
stellar spin rate $\Omega_s$ as a result of the magnetic torque:
\be
r^2B_zB_{\phi +}=-\dot M d(r^2\Omega)/dr,
\ee
For $\Omega_K(r_m)\gg\Omega_s$, we have $d(r^2\Omega)/dr\sim
r^2\Omega_K/\Delta r_m$, and
\be
r_m\simeq \left({\zeta\Delta r_m\over H}\right)^{2/7}
\left({\mu^4\over GM\dot M^2}\right)^{1/7}.
\label{eq:rm}\ee
Except for the first factor on the right-hand side, this is the usual
expression of magnetosphere radius for spherical accretion.

Note that equation (\ref{eq:rm}) applies to the case of $\Omega_K (r_m)\gg \Omega_s$,
or equivalently, the corotation radius $r_c$ 
(where $\Omega_K$ equals $\Omega_s$),
is much larger than $r_m$). For $\Omega_s\gg \Omega_K(r_m)$ (or $r_c<r_m$), 
we have
\be
r_m\simeq \left({\zeta\Delta r_m\over H}\right)^{1/5}
\left({\mu^2\over \dot M\Omega_s}\right)^{1/5}.
\label{eq:rm2}\ee
In practice, the difference between equations (\ref{eq:rm}) and
(\ref{eq:rm2}) is rather small, given the other uncertainties
in the problem.

The next question one can ask is whether $r_m$ (as determined above)
equals $r_{\rm in}$, the inner radius of the disk. There are two issues
to keep in mind:

(i) In principle, a disk-like structure, corotating with the star at
$\Omega_s$, could exist inside $r_m$. Note that this structure does not
Keplerian rotation, but is part of the corotating magnetosphere.  This ``disk''
plasma can drift inward by interchange instability (Spruit \& Taam
1990). At some point (depending on the field geometry), 
this ``disk'' can be disrupted and the plasma
flows along the field line to the magnetic polar cap of the star.

(ii) More importantly, the expression for $r_m$ given above assumes
steady state. In another word, $\dot M$ in the above expression is the local
value at $r_m$. It is possible that if mass accumulates at the boundary
(e.g., by the centrifugal barrier; see below), the local $\dot M$ is larger than
the mass supply rate ($\dot M_{\rm supply}$) at larger distance. This may lead to episodic
accretion. For example, when $\dot M$ is small, $r_m$ will be larger than
$r_c$, while $r_{\rm in}$ can be much less than $r_m$. Such episodic
accretion scenario has been explored by Spruit \& Taam (1993) and
D'Angelo \& Spruit (2010,2011,2012).
One issue concerning such a scenario is the possible non-axisymmetric
instabilities that may develop, allowing the accumulated mass to fall
rapidly through the magnetosphere onto the star.

\section{Star-Disk Linkage and Magnetic Field Inflation}
\label{sec-3}

\begin{figure}
\centering
\includegraphics[width=12.5cm,clip]{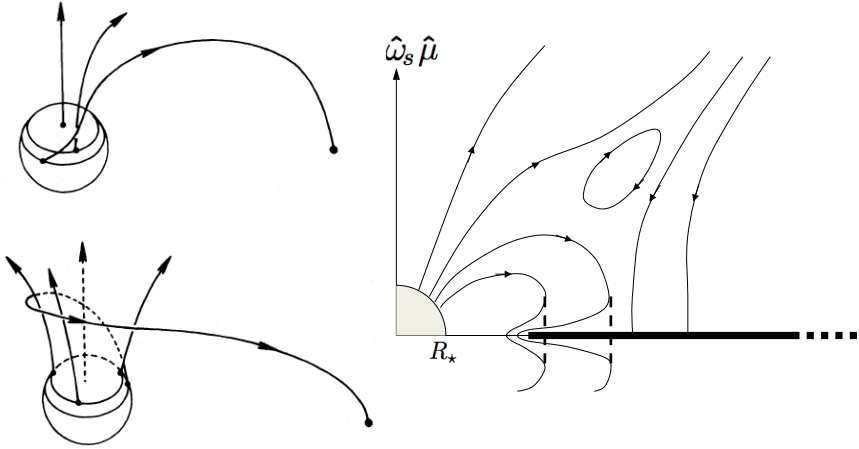}
\vskip -2.8truecm
\caption{Left: Magnetic field lines linking the star and disk are twisted
by differential rotation (Aly \& Kuijpers 1990). Right: Quasi-cyclic behavior
(see text).}
\label{f1}       
\end{figure}

When the stellar field lines penetrate some region of the disk, they
provide a linkage between the star and the disk. These field lines are twisted
by differential rotation between the stellar rotation $\Omega_s$ and the disk
rotation $\Omega(r)$, generating toroidal field. In steady state 
-- if it can be reached, the twisting is balanced by dissipation across
the disk. This gives the toroidal field just above the disk,
$B_{\phi+}\sim B_z (\Omega-\Omega_s)H^2/\eta$. With $\eta\sim \alpha Hc_s$,
we find $B_{\phi+}/B_z\simeq (\Omega-\Omega_s)/(\alpha\Omega)$. Obviously, except
for a narrow region around the corotation radius $r_c$, the toroidal field
is much stronger than the poloidal field. 

However, when the toroidal field becomes comparable to the poloidal
field, the flux tube connecting the star and the disk will start
expanding. This field inflation is driven by the pressure associated
with the toroidal field. As the fields open up, the star-disk linkage
is broken. Such field-opening behavior has been well-established
through theoretical studies and numerical simulations
in the contexts of solar flares and accretion disks (e.g., Aly 1985; Aly \& 
Kuijpers 1990; van Ballegooijen 1994; Lynden-Bell \& Boily 1994; 
Lovelace et al. 1995; Uzdensky et al. 2002; see Fig.~2). Thus, the maximum
toroidal twist is $|B_{\phi+}/B_z|_{\rm max}\sim 1$.

Given this constraint on the toroidal twist, steady-state disk-star
linkage is possible only very near corotation ($\Delta r/r_{\rm
  co}\sim\alpha$).  In general, we should expect a quasi-cyclic
behavior, involving several stages: (1) The stellar field penetrates
the inner region of the disk; (2) The linked field lines are twisted;
(3) The resulting toroidal fields drive field inflation; (4)
Reconnection of the inflated field restores the linkage.  The whole
cycle then repeats (see Aly \& Juijpers 1990; Uzdensky et al.~2002).

Given that such a quasi-cyclic behavior is inevitable, questions arise
as to how this behavior may manifest observationally. In particular,
is there any connection with QPOs in low-mass X-ray binaries 
(see van der Klis 2005 for a review) and other systems? In particular,
QPOs with frequencies around kHz have been observed in a number of accreting 
millisecond pulsars. The quasi-cyclic timescale is comparable to the
kHz QPO timescales, but exactly how these QPOs are produced remains 
unclear.

The quasi-cyclic state may also give rise to episodic outflows and winds,
to be discussed later.

\section{Torque on the Star, Propeller, Spin Equilibrium}
\label{sec-4}

Although a precise steady-state of the star-disk linkage cannot be
achieved, one can nevertheless ask the question: On average (i.e.,
over many dynamical times), what is the width ($\delta r$) of the
magnetically linked region in the disk? The reason that this
is of interest is that the width $\delta r$ directly affects the long-term
(secular) torque on the star and therefore its equilibrium rotation rate.

\begin{figure}
\centering
\includegraphics[width=12.5cm,clip]{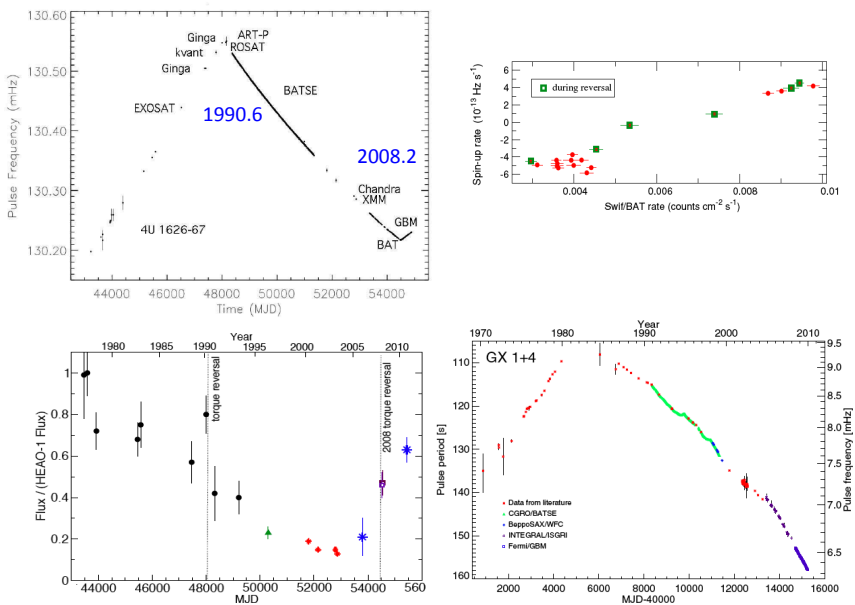}
\vskip -1.truecm
\caption{Left panels and the right-top panel: Spin evolution of 4U1626-67
and its correlation with the x-ray flux (see Camero-Arranz et al.~2010,2012).
Right-bottom panel: Spin evolution of GX 1+4 (see Chakrabarty et al.~1997; 
Gonzales-Galan et al.~2012).}
\label{f1}       
\end{figure}

In general, the torque in the star has two contributions, one associated with mass
accretion and another related to star-disk coupling:
\be
{dJ_\star\over dt}={\dot M}_{\rm acc} r_m^2\Omega(r_m)+T_m,
\ee
where $\dot M_{\rm acc}$ is the mass accretion rate onto the star,
which, in general, could be different from $\dot M_\infty$ or even
$\dot M$ (see above). The magnetic torque is
\be
T_m=\int_{r_m}^{r_m+\Delta r}dr\,2\pi r\left(-r{B_zB_{\phi+}\over 2\pi}
\right)
\simeq \zeta (r^2B_z^2)_{r_m}\Delta r,
\ee
where in the second equality we have assumed $\Delta r$ is small and
used $B_{\phi+}=-\zeta B_z$. Using $B_z\simeq (\mu/r^3)(r/H)^{1/2}$, we find
\be
T_m\simeq \left({\zeta \Delta r\over H}\right){\mu^2\over r_m^3}.
\ee
Using equation (\ref{eq:rm}) and assuming $\delta r$ is the same as the width of
the boundary layer $\delta r_m$, we find
\be
|T_m|\simeq \dot M\sqrt{GMr_m}.
\ee
Thus when $\dot M\simeq \dot M_{\rm acc}$, $|T_m|$ os of the same order of
magnitude as the accretion torque. Note, however, that $T_m$ can have
either positive or negative sign: While we expect $|\zeta|\sim 1$, the sign of
$\zeta$ depends on the location of the interaction zone relative to the corotation
radius $r_c$. For $r\simeq r_m<r_c$, we have $\zeta>0$ and the magnetic torque on
the star is positive, while for $r_m>r_c$, we have $T_m<0$.

The torque on the star is affected by the propeller effect. When the
accreting plasma penetrates the magnetosphere and gets attached to the
stellar field lines, it corotates with the star. If $r_m\Omega_s$
exceeds the escape speed $v_{\rm esc}(r_m)= (2GM/r_m)^{1/2}$, or
equivalently $r_m>2^{1/3}r_c$, the plasma may be ejected due to centrifugal 
force. The torque on the star should be modified to 
\be
{dJ_\star\over dt}={\dot M}_{\rm acc} r_m^2\Omega(r_m)+T_m-\dot M_{\rm eject}
r_m^2\Omega_s, 
\ee
where $\dot M_{\rm eject}$ is the mass ejection rate.  
In steady-state, we may expect $\dot M_{\rm acc}+\dot M_{\rm eject}=\dot M$.  
Precisely how much mass is ejected would depends on details of
dissipation/heating near the magnetosphere boundary, magnetic field
topology, etc., If the plasma is not ejected, it may accumulate near
the magnetosphere boundary, forming a dead disk (D'Angelo \& Spruit
2010,2012).  Is there any accretion onto the star through interchange
instabilities? Determining $\dot M_{\rm acc}$ and $\dot M_{\rm eject}$
in the canonical propeller regime is an important problem.

Of course, the reason we are interested in the torque on the star is
because we want to know what physics determines the equilibrium spin of
accreting magnetic stars. This is relevant to young stars (T tauri stars), 
accreting millisecond pulsars as well as long-period pulsars.
In the standard picture, already developed in the 1970s and 1980s,
the equilibrium rotation equals, to within a factor of two, the Keplerian
rotation rate at $r_m$. That is, equilibrium is reached when 
$r_c\simeq r_m$. Although there may be various factors (accretion torque,
magnetic torque, propeller, wind/outflows, etc) that compete to
drive the star's rotation, all these generally lead to the similar
equilibrium condition ($r_c\sim r_m$).

More constraining to the theory of magnetic star-disk interaction
is the spinup/spindown behavior of accreting x-ray pulsars.
Many x-ray pulsars have been observed to exhibit changing spinup and 
spindown behaviors over timescale of years (Bildsten et al.~1997).
Figure 3 shows two particularly striking examples: 4U1626-67 is an accreting
pulsar with spin period 7.66~s. The clean spinup before 1990.6 was followed
by a clean spindown, and another spinup phase starting 2008.2. The 
spindown/spinup transition lasted 150 days. The spinup rate $\dot\Omega_s$
is positively correlated with the x-ray flux (Camero-Arranz et al.~2010,2012).
In GX 1+4, it was found that during the spindown phase, the magnitude of the
spindown torque is positively correlated with the x-ray flux, 
$-\dot\Omega_s\propto F_x^{0.030\pm 0.07}$ (Chakrabarty et al.~1997;
Gonzalez-Galan et al.~2012). This latter correlation is difficult to explain.
In fact, it directly contradicts the prediction of the original Ghosh-Lamb
model. Perhaps some variations of the models based on variable
$\dot M_{\rm acc}$ and $\dot M_{\rm eject}$ can do the job, but exactly 
how is not clear (see Locsei \& Melatos 2004; Perna, Bozzo \& Stella 2006;
Dai \& Li 2006 for related works).
Understanding this spindown/spinup behavior and its correlation with
the accretion rate remains an outstanding unsolved problem.

\section{Outflows}
\label{sec-5}

Outflows is a natural consequence of magnetosphere-disk interaction.
We have already touched upon two effects that may give rise to
outflows.  

(1) The quasi-cyclic star-disk linkage (see Section 3) and
associated field inflation inevitably lead to ejection of plasma from
the magnetosphere boundary.  This episodic outflows have been seen in
simulations (see Romanova et al. 2009; Zanni \& Ferreira 2013), but may also be
related to the X-wind model of Shu et al. (although the ``canonical''
X-wind model was developed for the case when the star has already reached
spin equilibrium, thus $r_m\simeq r_c$, and the wind is steady).

(2) In the propeller regime, plasma may be ejected from the magnetosphere.
Such outflow is distinct from the boundary outflow discussed in (1).
It maybe more more collimated and has a faster speed (see Lii et al.~2013
for recent simulations).

In addition to the above two outflow launching mechanisms, there are two other
possibilities:

(3) Accretion-induced stellar wind (see paper by R. Pudritz).

(4) Outflow from the disk through magneto-centrifugal mechanism
a la Blandford-Payne. Perhaps such outflow is less collimated and
has a slow velocity.

It is possible that some of these mechanisms are related to each other and
therefore not distinguishable in practice (both in simulations and observations).
Obviously, it is important to confront all these mechanisms with observations.
Outflows/jets from young stellar objects are well known and discussed elsewhere. 
But there are also observed in accreting neutron stars in binaries. Perhaps
the famous is the relativistic jet from Circinus X-1 (Fender et al.~2004;
Armstrong et al.~2013),
a bright and highly variable x-ray binary containing a neutron star 
on an eccentric orbit (17 days) with a normal star companion.
It is interesting to note that jets and outflows have been observed
in different types of neutron star binaries, including Atolls sources 
(such as Aql X-1), Z-sources (Sco X-1) and accreting millisecond
x-ray pulsars (SAX J1808). It is usually thought that the key difference
in these sources lies in their magnetic field strengths, with accreting pulsars
having the highest fields ($\sim 10^9$~G).

It is also of interest to compare the jet phenomenology in black-hole x-ray
binaries and neutron-star x-ray binaries. In the case of black hole, compact steady
jets are produced in the low-hard state, when the innermost disk is 
replaced by a geometrically thick flow/corona; episodic/ballistic jets are
produced in the transitional state between the low state and the high, thermal
state; no jets are produced in the ``clean'' thermal state. By contrast,
for neutron star systems, jets appear to be produced in the soft x-ray thermal
(called ``banana'') states (for example, Ser X-1, 4U1820-30; see
Migliari et al. 2011). The same can be said of young stars, which are always
in the ``thermal disk'' state. Perhaps this difference points to the important
role of magnetosphere, which is absent in black-hole systems.

\section{Misaligned Stellar Fields}
\label{sec-6}

The physics and issues we have discussed so far are applicable
regardless of whether the stellar magnetic dipole is aligned or misaligned
with the stellar spin axis. In general, the stellar dipole axis is misaligned
with the spin (this is required for accreting x-ray pulsars). Are there any new
effects? The answer is yes. One obvious effect is that with misaligned magnetic
fields, the plasma in the magnetosphere can flow to the polar cap more
easily (the so-called ``funnel flow''), thereby producing pulsed emission 
from the stellar surface. With a very aligned stellar field, this may be difficult,
and magnetosphere accretion may proceed through instabilities, so that the
plasma may land on the stellar surface more in the equatorial region than 
the polar caps. This obviously will have different observational consequences
(see M. Romanova's paper in this proceedings).

\begin{figure}
\centering
\includegraphics[width=12.5cm,clip]{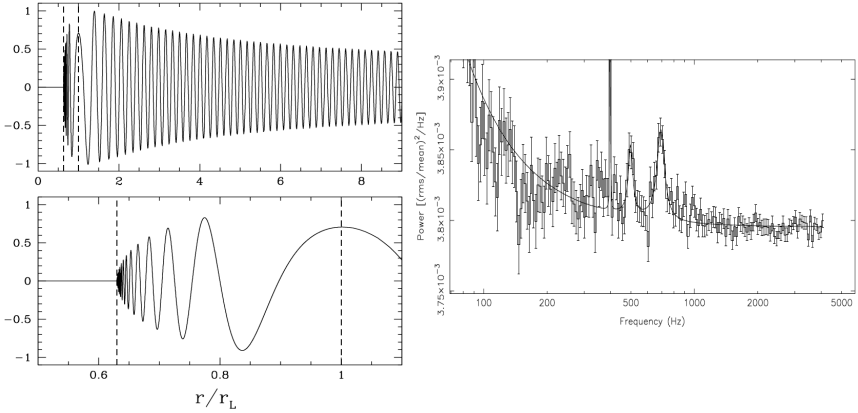}
\vskip -3.6truecm
\caption{Left panel: Bending waves in the disk excited by magnetic force from a
misaligned stellar dipole. The wave is most visible at the Lindblad/vertical
resonance (from Lai \& Zhang 2008).
Right panel: Power density spectrum of accreting millisecond pulsar 
SAX J1808.4-3658, showing the spin frequency $\nu_s=401$~Hz and two
kHz QPOs, with $\nu_h-\nu_l\simeq \nu_2/2$ (from van der Klis 2005).}
\label{f4}       
\end{figure}

Another effect of misaligned stellar dipole is that it can excite
non-axisymmetric waves in the disk through magnetic forcing (Lai \&
Zhang 2008; see Romanova et al.~2013 for recent simulations).
For example, the vertical magnetic force on the disk can
be written as
\be
F_z(r,\phi,t)=F_{\omega}(r)\exp(im\phi-i\omega t),
\ee
where $m=1$. The forcing frequency $\omega$ can be either $\Omega_s$ or
$2\Omega_s$. While $\omega=\Omega_s$ is obvious, the $2\Omega_s$ component 
requires some explanation. It arises because while the dipole field
at the given point in the disk varies as $\cos \Omega_s t$, the induced
screening current in the disk (by virtue of its high conductivity) also varies
as $\cos \Omega_s t$; thus the force has a term proportional $\cos^2\Omega_s t$.
(Note that this term is present only when the disk is misaligned with the
stellar spin; see below). This periodic vertical force will excite bending waves
in the disk. These bending waves are launched at the Lindblad/vertical
resonance, where the Doppler-shifted forcing frequency
$\omega-\Omega(r)$ matches the intrinsic disk vertical epicyclic
frequency $\Omega_z\simeq \Omega$.
Thus, the perturbation is most ``visible'' at $r=r_L$ given by
$\Omega(r_L)=\omega/2=\Omega_s/2,\Omega_s$. 

These magnetically excited bending waves may have interesting
applications to kHz QPOs observed in accreting millisecond pulsars
(van der Klis 2005). In particular, in SAX J1808.4-3658 (with the spin
frequency $\nu_s=401$~Hz), the two kHz QPOs frequencies both vary as
the source luminosity changes, but the difference frequency
$\nu_h-\nu_l$ remains constant and equals $\nu_s/2$ to within a few
Hz. In XTE J1807.4-294 (with $\nu_s=191$~Hz), the difference $\nu_h-\nu_l
\simeq \nu_s$. It seems that these are not pure coincidences.
(One should add that for most LMXBs which do not show coherent pulsations, 
the difference frequency of kHz QPOs do not exhibit similar correlation with
the neutron star spin.) Perhaps the beating of high-frequency
QPO ($\nu_h$) generated at the magnetosphere (or inside) with the perturbed
disk plasma at the Lindblad/vertical resonance can explain 
these intriguing observations.

\section{Stellar Spin - Disk Misalignment}
\label{sec-7}

In the standard picture magnetic star-disk interaction, it is usually
assumed that the stellar spin axis is aligned with the disk axis (the
disk normal vector). This seems reasonable since the star may have
gained substantial angular momentum from the accreting gas in the
disk.  However, magnetic interaction between the star and the inner
region of the disk may (if not always) change this simple picture
(Lai 1999; Lai et al.~2011).

\begin{figure}
\centering
\includegraphics[width=12.5cm,clip]{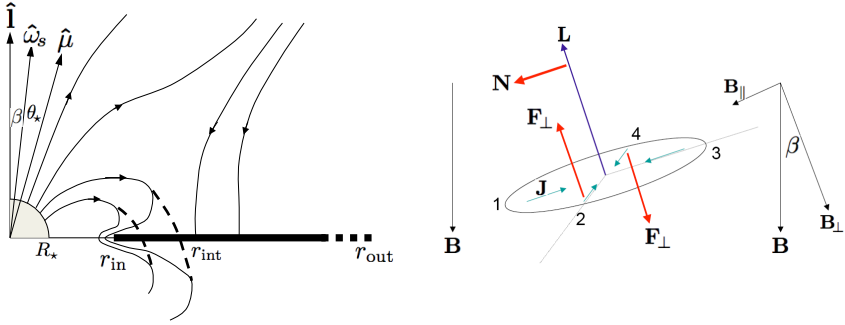}
\vskip -5.0truecm
\caption{Left panel: A sketch of magnetic field configuration in a
  star-disk system for nonzero $\beta$ (the angle between the disk
  axis and the stellar spin axis) and $\theta_\star$ (the angle
  between the stellar dipole axis and the spin axis). Part of the
  stellar magnetic fields (dashed lines) penetrate the disk in the
  interaction zone between the disc inner radius $r_{\rm in}$ and
  $r_{\rm int}$ in a cyclic manner, while other field lines are
  screened out of the disc. The closed field lines are twisted by the
  differential rotation between the star and the disc, which leads to
  a magnetic braking torque and a warping torque. The screening
  current in the disk leads to a precessional torque.  Right panel:
  A toy model for understanding the origin of the warping torque.  A
  tilted rotating metal plate (with angular momentum ${\bf L}$) in an
  external magnetic field ${\bf B}$ experiences a vertical magnetic
  force around region 2 and 4 due to the interaction between the
  induced current ${\bf J}$ and the external ${\bf B}_\parallel$,
  resulting in a torque ${\bf N}$ which further increases the tilt
  angle $\beta$.}
\label{f5}       
\end{figure}

Before being disrupted at $r_{\rm in}$ by the stellar magnetic field, 
the disk generally experiences nontrivial
magnetic torques from the star.  These torques are of two types: (i) A
warping torque ${\bf N}_w$ which acts in a small interaction region
$r_{\rm in} < r < r_{\rm int}$ (with width $\Delta r$),
where some of the stellar field lines are
linked to the disk.  These field lines are twisted by the differential
rotation between the star and the disk, generating a toroidal field
$\Delta B_\phi=\mp \zeta B_z^{(s)}$ from the quasi-static vertical
field $B_z^{(s)}$ threading the disk, where $\zeta\sim 1$ and the
upper/lower sign refers to the value above/below the disk plane. Since
the toroidal field from the stellar dipole $B_\phi^{(\mu)}$ is
symmetric with respect to the disk plane, the net toroidal field
differs above and below the disk plane, giving rise to a vertical
force on the disk. While the mean force (averaging over the azimuthal
direction) is zero, the uneven distribution of the force induces a net
warping torque which tends to push the orientation of the disk angular
momentum $\hatl$ away from the stellar spin axis $\hat{\bOmega}_s$.
The essential physics of the warping torque can also be understood
from the ``laboratory'' toy model depicted in Fig.~5.  (ii) A
precession torque ${\bf N}_p$ which arises from the screening of the
azimuthal electric current induced in the highly conducting disk. This
results in a difference in the radial component of the net magnetic
field above and below the disk plane and therefore in a vertical force
on the disk.  The resulting precession torque tends to cause $\hatl$
to precess around $\hat\bOmega_s$.

The two magnetic torques (per unit area) on the disk can be written as
\be
{\bf N}_w = -(\Sigma r^2\Omega)\cos\beta\,
\Gamma_w \,\hatl\times(\hat{\bOmega}_s\times\hatl),\qquad
{\bf N}_p=(\Sigma r^2\Omega)\cos\beta\,
\Omega_p \,\hat{\bOmega}_s\times\hatl,
\label{eq:torque}\ee
where $\Sigma(r)$ is the surface density, $\Omega(r)$ the
rotation rate of the disk, and $\beta(r)$ is the disc tilt angle (the angle
between $\hatl (r)$ and the spin axis $\hat\bOmega_s$).
The warping rate and precession angular frequency at radius $r$ are
given by
\be
\Gamma_w (r)=\frac{\zeta\mu^2}{4\pi r^7\Omega(r)\Sigma(r)}\cos^2\theta_\star,
\qquad
\Omega_p (r)=-\frac{\mu^2}{\pi^2 r^7\Omega(r)\Sigma(r)
  D(r)}\sin^2\theta_\star,
\label{eqn:Omega_w}
\ee
where $\theta_\star$ is the angle between the magnetic dipole axis and the
spin axis, and the dimensionless function $D(r)$ is somewhat less than unity.

\subsection{Applications to QPOs/Variabilities}

The magnetically driven warping and precession effects discussed
above may be relevant to the low-frequency (10-50~Hz)
QPOs (the so-called horizontal branch oscillations)
observed in neutron star LMXBs. Stella \& Vietri (1998) suggested that LFQPOs
are associated with the Lense-Thirring precession of the inner
accretion disk around the rotating neutron star. If the LFQPO and
the kHz QPOs are generated at the same special radius in
the disk, this implies a quadratic relation between the
LFQPO frequency and the kHz QPO frequency, in rough
agreement with observations (usually for a certain range of
X-ray fluxes). However, in order to have precession, the inner disk
must be tilted relative to the stellar spin. The magnetic effects may induce
the required disk warp and contribute the the disk precession
(Shirakawa \& Lai 2002a, Pfeiffer \& Lai 2004).

Magnetically driven warping and precession may also explain
millisecond variabilities in high-field accreting X-ray pulsars 
(Shirakawa \& Lai 2002b) as well as some photometric variabilities observed in
T Tauri stars (such as AA Tau; see Terquem \& Papaloizou 2000;
Bouvier et al.~2013).

\subsection{Possible Connection to (Exo)Planetary Systems}

One of the recent surprises in exoplanetary sciences is that many hot
Jupiter systems (orbital periods of order a few days) are found to
have a misaligned orbital axis relative the spin axis of the host
star.  Among the several dozens of hot Jupiter systems with
measurements the stellar obliquity, about $60\%$ have an orbital axis
aligned (in sky projection) with the stellar spin, while the other
systems show a significant spin-orbit misalignment, including some
with retrograde orbits.

In our own Solar System, except for Pluto, all planets
outside 1~AU lie within 2$^\circ$ of the ecliptic plane, while the
Sun's equatorial plane is inclined by 7$^\circ$ with respect to the
ecliptic. It is not clear whether the difference between 2$^\circ$ and
7$^\circ$ is significant or needs an explanation.

The observed spin-orbit misalignment in many hot Jupiter systems
indicates that gravitational interactions 
between multiple planets and/or secular interactions with a distant
planet or stellar companion play an important role 
in shaping the architecture of planetary systems.
Nevertheless, there is an alternative possibility: misaligned
protostar and protoplanetary disk. Such misaligned disk may explain the
$7^\circ$ ``anomaly'' in Solar System and affect the initial condition 
of planetary formation.

As discussed above, the inner region of the disk around a magnetic protostar 
experiences a warping torque and a precessional torque.
If we imagine dividing
the disk into many rings, and if each ring were allowed to behave
independent of each other, it would be driven toward a perpendicular state
and precess around the spin axis
of the central star. Obviously, real protoplanetary disks do not
behave as a collection of non-interacting rings: Hydrodynamic stresses
(bending waves and viscosity) provide strong couplings between different rings,
and the disk will resist the inner disk warping. Detailed calculations
by Foucart \& Lai (2011) showed that for most reasonable stellar/disk
parameters, the steady-state disk warp is rather small because of
efficient viscous damping or propagation of bending waves. Thus the inner
disk direction is approximate the same as the outer disk direction.

What is happening to the stellar spin? Is there a secular change to
the stellar spin direction? The answer is maybe. The back-reaction
magnetic torque from the disk on the stellar spin can push the stellar
spin axis away from alignment. The magnetic misalignment
torque on the star is of the same order as the fiducial accretion torque
$\dot M\sqrt{GM_\star r_{\rm in}}$. Assuming the spin angular momentum
$J_s=0.2M_\star R_\star^2\Omega_s$, we find the spin evolution time to be
\be
t_{\rm spin}=(1.25\,{\rm Myr})\left(\!{M_\star\over 1\,M_\odot}\!\right)
\!\!\left({{\dot M}\over 10^{-8}{M_\odot}{\rm yr}^{-1}}\right)^{\!-1}
\left(\!{r_{\rm in}\over 4R_\star}\!\right)^{\!-2}
\!\!{\Omega_s\over\Omega(r_{\rm in})}.
\label{eq:tspin}\ee
Given enough time, the stellar spin axis will evolve towards the
perpendicular state and even the retrograde state.  Unfortunately,
there are other torques acting on the star which counter-act the
warping torque. In particular, the accretion torque tends to align the
stellar spin axis with the disk. Since the magnetic torque and the
accretion torque are of the same order of magnitude, we cannot be
make a definitive conclusion. All we can say is that both alignment
and misalignment between the stellar spin axis and the disk axis are
possible (see Lai, Foucart \& Lin 2011 and Foucart \& Lai 2011 for more details).

{\bf Acknowledgments:}
I thank the organizers for the invitation and support for me 
to attend the conference on the Physics at the Magnetosphere Boundary
in Geneva. This work has been supported in part by NSF grants
AST-1008245, 1211061 and NASA grant NNX12AF85G.




\end{document}